\documentstyle[12pt]{article}

\input psfig.sty

\newcommand{\be}{\begin{equation}}
\newcommand{\ee}{\end{equation}}
\newcommand{\bees}{\begin{eqnarray}}
\newcommand{\ees}{\end{eqnarray}}
\newcommand{\gsim}{\lower.7ex\hbox{$\;\stackrel{\textstyle>}{\sim}\;$}}
\newcommand{\lsim}{\lower.7ex\hbox{$\;\stackrel{\textstyle<}{\sim}\;$}}
\newcommand{\ra}{\rightarrow}
\newcommand{\pa}{\partial}
\newcommand{\mpl}{m_{\rm Pl}^{(11)}}

\begin{document}

\baselineskip=17pt
\pagestyle{plain}
\setcounter{page}{1}

\begin{titlepage}

\begin{flushright}
CERN-TH/98-361\\
IFUP-TH/52-98\\

\end{flushright}
\vspace{5 mm}

\begin{center}
{\LARGE\bf D-branes and Cosmology}

\end{center}

\vspace{4mm}

\begin{center}
{\large  Michele Maggiore$^{a,}$\footnote{
maggiore@mailbox.difi.unipi.it} and 
Antonio Riotto$^{b,}$\footnote{riotto@nxth04.cern.ch. On leave of absence
from the University of Oxford, Theory Group, Oxford, UK.} }\\
\vspace{1cm}
{\em $^a$ INFN and Dipartimento di Fisica, via Buonarroti 2, I-56100
Pisa, Italy}\\
\vspace{0.3cm}

{\em $^b$ Theory Division, CERN, CH-1211, Geneva 23, Switzerland}

\end{center}
\vspace{2mm}
\begin{center}
{\large Abstract}
\end{center}
\noindent

D-branes, topological defects in string theory on which string
endpoints can live, may  give new insight into the understanding
of the cosmological evolution
of the Universe at early epochs.
We analyze the 
dynamics of D-branes in curved backgrounds and
discuss the parameter space of M-theory as a function of the coupling
constant and of the curvature of the Universe. We show that D-branes may be  
efficiently produced by gravitational effects. Furthermore,
in curved spacetimes the transverse fluctuations of the D-branes
develop a tachyonic mode and when the fluctuations grow larger than the
horizon  the branes
become tensionless and break up. This signals a transition to a new regime.
We discuss the  implications of our findings  for
the singularity problem present in string cosmology, suggesting the  
existence of a
limiting value for the curvature which is in agreement with the value
suggested by the
cosmological version of the holography principle. We also comment on 
possible implications for the so-called  brane world scenario, 
where the Standard Model gauge
and matter fields live inside some branes while gravitons live in the bulk.

\end{titlepage}
\newpage

\baselineskip=24pt

\section{Introduction}
Recent important developements both in field theory~\cite{SW} and
in string theory~\cite{Dbrane}  are based on the idea that in different
regions of the parameter space there are  different 
light modes and therefore different
effective actions. 
A famous example is electric-magnetic duality: at weak
coupling the fundamental degrees of
freedom are electrically charged fields 
while  monopoles are heavy. At some
point of the moduli space, however, magnetic monopoles become the
only light states of the theory.
Switching to an effective description in terms of the new light fields
allows to address problems that, from the point of view of the original
formulation, are  difficult and non-perturbative. 

Identifying properly the light modes is especially important when
dealing with the problem of singularities. An important example 
is the conifold singularity that appears in Calabi-Yau
compactifications when there is a non-trivial 3-cycle with period
$z$. In the limit $z\ra 0$ the moduli space apparently develops a
singularity. However, it was realized in~\cite{Str} that a D3-brane
wrapping on this cycle has a mass $\sim |z|$, and becomes massless in
the limit $z\ra 0$. The
singularity is due to the fact that we have integrated out this
massless state. Including this state in the low energy
effective theory shows that there is no singularity (see also~\cite{OV}).

A similar situation may appear when  dealing with the issue of cosmological
singularities  in string theory. In general, in any
cosmological model based on string theory, one has to face  a regime of
strong coupling and large curvatures when approaching the big-bang
singularity. Consider,  for instance,  the
cosmological evolution obtained from the  lowest order effective action
of string theory, taking as initial conditions  a flat, weakly
coupled configuration~\cite{GV}. The classical solution
evolves toward a singularity in the strong coupling/large curvature
regime. Most of the approaches adopted to smear out this 
singularity  consist in  adding
perturbative $\alpha '$ or loop corrections to the string
effective action~\cite{corr1,corr2}. It is  possible that these
corrections regularize the classical solutions -- hopefully before
entering the  fully strongly coupled regime --   thus providing a consistent
and satisfying solution of the singularity problem. 

If instead the evolution enters  the full strongly
coupled/large curvature regime, one can ask whether it is possible
to turn to a new description, which  amounts to summing
up the whole tower of massive closed string modes and/or to resum 
non-perturbatively the loop expansion.
To get some insight into  this question, it is 
instructive to consider an analogous situation in more details.
Let us consider  two parallel Dirichelet
$p$-branes.
The amplitude of the interaction  is given by~\cite{Pol}
\be\label{A}
{\cal A}=V_p\int\frac{d^{p+1}k}{(2\pi )^{p+1}}
\int_0^{\infty}\frac{dt}{2t}\sum_{I}e^{-2\pi\alpha ' t(k^2+m_I^2)},
\ee
where the sum $I$ is over the spectrum of an open string stretching
between the two branes and $V_p$ is the brane volume. 
This amplitude is reproduced by an effective
potential ${\cal V}(v,r)$, 
which depends on the D-branes separation $r$ and
relative velocity $v$, since $m_I=m_I(r,v)$. Expanding in powers of $v$,
${\cal V}(v,r)=\sum_n v^n{\cal V}_n(r)$, one finds~\cite{DKPS} 
\be\label{V}
{\cal V}_n(r)\sim\int_0^{\infty}\frac{dt}{t^{(3+p)/2}}g_n(t)
e^{-r^2t/(2\pi \alpha ')}
\ee
where $g_n(t)$ is non-singular for finite $t$ and behaves like $t^n$ 
as $t\ra \infty$. As emphasized in
Ref.~\cite{DKPS}, the  singular behaviour
 at $r\ra 0$ may only come from the $t\ra \infty$ integration
region, and eq.~(\ref{A}) shows that this region is
dominated by the light modes of the open string stretching between
the two branes. Therefore, the
short distance region can be studied truncating eq.~(\ref{A}) to the
first few open string modes, and the short distance 
singularity is controlled by the
lightest mode.

However -- exchanging the world-sheet space and time -- 
the  amplitude ${\cal A}$ can also be seen as due to closed string
exchange, and the poles from the massless modes of the closed string
come from the
integration region $t\ra 0$  \cite{Pol}. This
limit determines the large $r$ behaviour of the interaction.
Therefore, the amplitude can be computed exactly  either as a sum over
all closed string modes  or as a sum over all open string modes, but the
truncation to the first few modes of the open and closed string
is valid in different domains~\cite{DKPS}: a
truncation to the lightest modes of the open string is appropriate in
the limit $r\ll \lambda_s$, where $\lambda_s$ is the string length, and it is
therefore suitable for the study of  short distance singularities;  the
truncation to the lightest modes of the closed string is appropriate
in the limit $r\gg \lambda_s$, where it reproduces the supergravity
results.

Similarly -- if we consider the D-brane as a solitonic solution in 
supergravity -- the metric and the dilaton field $\phi$ diverge  when 
approaching the
D-brane, {\it i.e.}
 as the distance from the brane goes to zero. The considerations presented 
above suggest that  
 this divergence is in fact not meaningful,
since   supergravity is not  the right tool  in such a  regime.
Trying to cure this divergence by including the pile-up of massive closed 
string states in the supergravity effective action -- {\it i.e.}  trying   
 to catch the features of the region $t\ra \infty$ performing a perturbative
expansion around $t=0$ -- is not the most  economic procedure. One should 
instead adopt  the more appropriate
open string picture. The effective Lagrangian which is suitable  in this
regime is written in terms of 
the lightest mode of the stretched string, with mass $m\sim r/\lambda_s^2$. 
In other words, one should not look for a
regular metric and dilaton field. The appropriate description is now in
terms of different low energy fields.

The above discussion  suggests the following approach to cosmological 
singularities. Rather than
trying to obtain a non-singular evolution for the metric and the
dilaton, close the singularity
one should describe the system in terms of  new low energy
modes  appropriate to the strong coupling/large curvature  regime,
and  understand the matching between the standard
10D-supergravity and the new effective Lagrangian description
(related ideas have been presented in ref.~\cite{LW}).
In the  strong coupling regime, 
D-branes are the fundamental players.  In the string frame their tension 
is $\sim 1/g=e^{-\phi}$ (in string units) while the tension of 
the fundamental string is
${\cal O}(1)$. So at weak coupling D-branes are heavy and irrelevant to the
dynamics, while at strong coupling they are the light objects of the
theory. 
In particular the D0-brane of type IIA has a mass
\begin{equation}
m=\frac{1}{g\sqrt{\alpha '}}
\end{equation}
 and the D-string of type IIB has a
string  tension $1/(g 2\pi\alpha ')$, while the F(undamental)-string has
tension $2\pi\alpha '$. The duality between D-string and F-string is
even more clear using the Einstein frame, where the F-string has a
tension $g^{1/2}/(2\pi\alpha ')$ and the D-string
$g^{-1/2}/(2\pi\alpha ')$.
D-branes also naturally lead us to think in terms of 11 dimensions. For
instance, the D0 branes of 
type IIA string  theory from the 11-dimensional
point of view have a very simple interpretation as Kaluza-Klein (KK)
modes of M-theory compactified on $S^1$~\cite{matrix} which -- 
at low energy and large distances -- is described by 
11D-supergravity. D-branes are therefore expected to play a
fundamental role in the very early epochs of the cosmological
evolution, if the evolution goes through a strongly coupled regime.

The coupling $g$ is not the only parameter that determines which objects
are light and which are heavy. In a curved background the
characteristic value of the curvature is crucial as well. 
We will show that
the background gravitational field modifies
the D0-brane mass and the D$p$-branes tension.
A first interesting phenomenon takes place when the expansion
rate of the Universe $H$ becomes of order of the inverse radius of the
11$^{th}$ dimension $1/R_{11}$. At this stage the 11$^{th}$ dimension
opens up, D0-branes become tachyonic and are efficiently generated.
A second critical phenomenon occurs when $H$ is of the order of the
inverse of the 11$^{th}$-dimensional Planck length $\ell_p$. The
renormalization of the brane tension due to the gravitational field
becomes crucial and D-branes become tensionless. 
The would-be Goldstone bosons of the broken translation invariance become
tachyonic and there appear large transverse fluctuations.
At this stage the cosmological evolution cannot be described by the
low-energy string effective action and one has to take into account
the appearance of new light degrees of freedom such as the
pseudo-Goldstone bosons and new phenomena like the breaking up and
continous creation of D-branes.
Our results indicate that the value of the curvature cannot
exceed  a limiting value of order  of $\ell_p^{-2}$. Interestingly
enough, this bound turns out to be in agreement both with the value
suggested by the cosmological version of the holographic principle.

The paper is organized as follows.
Taking  the Hubble parameter $H$ as an indicator of the curvature,
the first step of our strategy amounts to identifying what is the
most appropriate effective lagrangian description in different regions
of the $(H,g)$ plane. This will be the subject of section~2. 
In section~3 we discuss the crossover into the 11D region and discuss 
the production of D0-branes. 
In section~4 we compute the renormalization of the D$p$-brane tension
and study the transition to a D-brane dominated regime.
In section~5 we present our conclusions and discuss the implications 
of our findings for pre-big-bang cosmology and for the recent developements
on the possibility of lowering the string scale down to a TeV.

\vskip 0.5cm 

\section{The cosmological `phase diagram' of M-theory }

\vskip 0.5cm

Let us first  discuss  the parameter space of M-theory
compactified on $S^1$, as a function of the coupling $g$ and of the
typical value of the curvature. In the following we will  take the Hubble
parameter $H$ (or $|H|$ for a contracting Universe)
as the measure of the characteristic value of the energy and $H^2$ as the 
characteristic value of the curvature. Our immediate goal is to understand
what is the most appropriate description in each region of the $(H,g)$
plane: in different regimes for the
energy and the coupling we expect different effective descriptions,
thus defining a  `phase diagram' of the theory.
We will then discuss  the crossover between different regions.

Already  inspecting the limit $g\gg 1 $ at fixed $H$, we  realize that
the phase diagram  must have a non-trivial structure. In fact,
when $g$ gets larger and larger, the tension of all D-branes become small, 
and therefore the mass of the D0-brane, 
the mass of the tower of excitations of the D-string, {\it etc.} all 
become very small. Since at this point we have (a plethora of) 
new light states, the
correct effective action at large $g$ has nothing to do with the 
10-dimensional supergravity
Lagrangian, even if one includes string loop corrections. The light modes are
simply different.

 Similarly -- as we will show in section~\ref{5} -- 
something even more dramatic may happen:
when  $H$ increases at {\it fixed}  coupling {\it all} the D$p$-branes 
become tensionless!  This is telling us that
in this region of the parameter space the important degrees of freedom
are D$p$-branes.

In  fig.~1 we present  
the cosmological version of the  phase diagram for M-theory compactified
on $S^1$ \cite{GHV}. The radius of the $11^{th}$ dimension is
$R_{11}=g\lambda_s$ and the first Kaluza-Klein mode has a mass
$1/R_{11}$. If the typical energy scale $\sim H\ll 1/R_{11}$  the
theory is effectively 10-dimensional, while at $H\gsim 1/R_{11}$ the $11^{th}$
dimension becomes accessible. The solid curve $\lambda_sH=g^{-1}$ therefore
separates the region where the effective theory is 10-dimensional from
the region where it is 11-dimensional.

\begin{figure}[tc]
\centering
\psfig{file=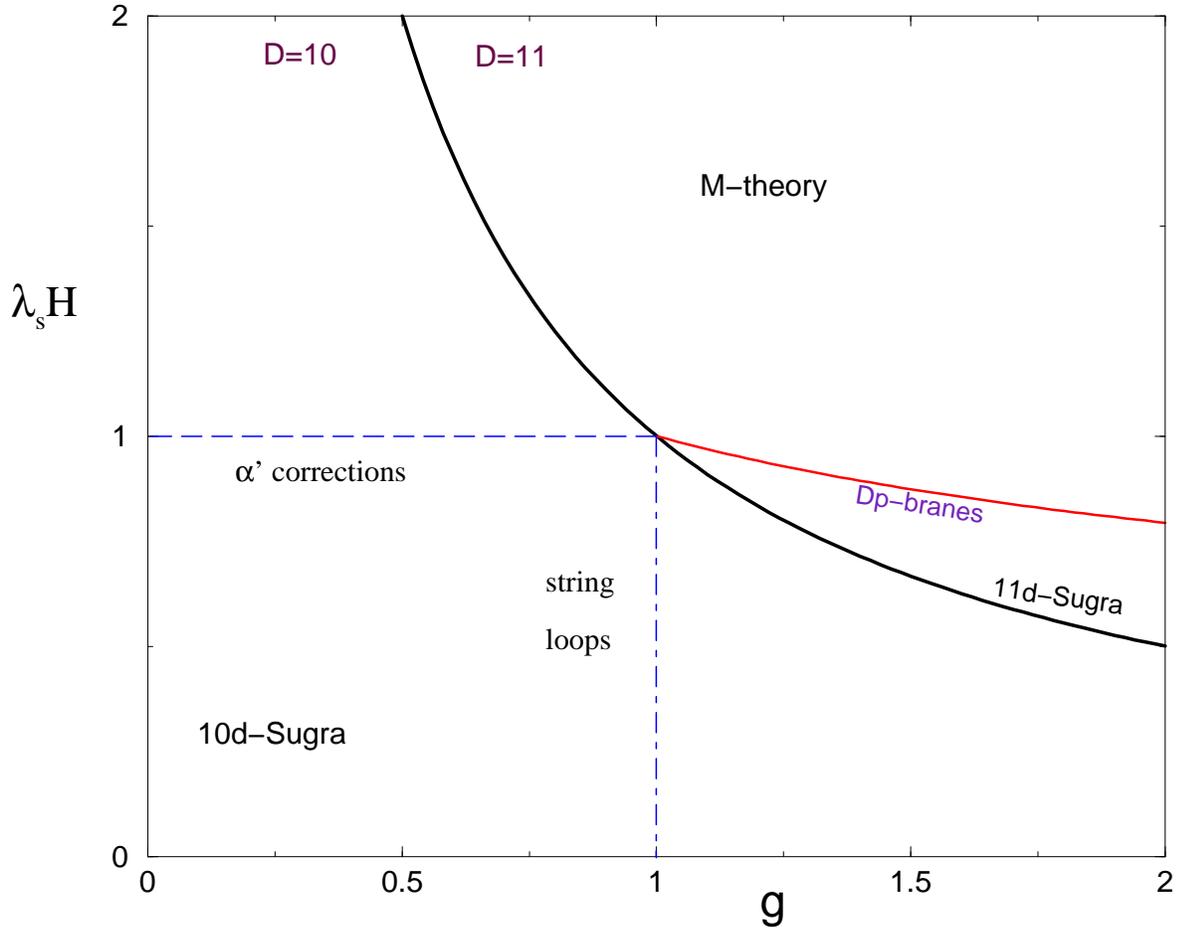,width=\linewidth,angle=270}
\caption{The phase diagram for M-theory compactified on $S^1$.}
\end{figure}

On the 10-dimensional side, the region near the origin corresponds to
low curvatures and weak couplings, and here the description in terms of the
lowest order closed string effective action is adequate. String theory
provides two kind of corrections to this 10D supergravity action:
$\alpha '$ corrections, {\it i.e.} corrections associated with higher powers
of the Riemann tensor, which become important as
$\lambda_sH$ approaches one, and string loop corrections, which become
important as $g$ approaches one. The dashed lines in fig.~1 limit
the validity of the perturbative expansion of the low-energy
10-dimensional string effective action.

On the 11-dimensional side, we can distinguish two regions separated
by the curve $\ell_{p}H=1$, where $\ell_p\sim
g^{1/3}\lambda_s$ is the 11-dimensional Planck length. The curve 
is indicated by the label  ``D$p$-branes". 
If we cross the border  between the 10D and the 11D regions at large
values of $g$ and we are just above the line identified by the relation
$\lambda_s H=g^{-1}$, 
 an adequate
description is still provided by 11-dimensional supergravity. 
In fact, now the large $g$ limit is under control
since it corresponds to the decompactification
limit $R_{11}\gg\lambda_s$, and at the same time $H\ll M_p$, 
where $M_p=1/\ell_p$ is the 11-dimensional Planck mass.
The 10D gravitons (to be identified with the 11-D gravitons 
with vanishing momentum along the 11$^{th}$ dimension)
are still the  relevant massless modes.  
Quantum production of D0-branes by the
gravitational field takes place and 
 a new  degree of freedom appears, which is the graviton with 
$p_{11}$ non vanishing, or equivalently a gas of weakly interacting
D0-branes.  As we will compute
in sect.~\ref{3}, this mode becomes tachyonic at the transition, signalling
the opening up of the new dimension.

This description will eventually break down if 
the system approaches the curve labelled ``D$p$-branes". If we approach
this curve at fixed $H$ and increasing $g$  more and more KK-modes
must be included
as light degrees of freedom and more and more D$p$-brane states 
become light because their tensions decrease as $g^{-1}$. 
However -- as we will show in section 5 -- something dramatic happens even 
if  we approach the curve
from below at fixed $g$ and increasing $H$:
{\it all} the D$p$-branes
become tensionless and they become the fundamental degrees of freedom! 
Therefore  the Universe enters a  regime of
D$p$-brane domination that leads the system to  a transition into the
full M-theory regime.

Thus, moving from the 10D Sugra region into the 11D Sugra region and
then toward the full M-theory regime, the basic modes are at first the
massless closed string modes, then both massless closed string modes
{\it and} D0-branes ({\it i.e.} 11-dimensional gravitons), 
and finally only D$p$-branes.

Of course, what is the  proper description in the full M-theory region 
is still an open   question (with matrix theory~\cite{matrix} being a very
interesting candidate), but in the following we will argue that its
knowledge is   not really necessary since our findings seem to
indicate that the  
line labelled ``D$p$-branes' in fig. 1 was never crossed during the 
evolution of the Universe.   

\section{D0-branes production and the opening of the $11^{th}$ dimension}
\label{3}

We now discuss the crossover between the 10D-Sugra and 11D-Sugra regime.
Let us begin by recalling a few well known facts about
the correspondence between 11-dimensional fields 
or branes compactified on $S^1$,
and 10-dimensional type IIA quantities~\cite{Dbrane}.
On the 10-dimensional side, the fields are the 10D metric
$g_{\mu\nu}$, the dilaton $\phi$,  the antisimmetric tensor
field $B_{\mu\nu}$ and  RR tensor fields, plus the
fermionic partners. On the 11-dimensional side, the
fields are the metric $g_{MN}$, with $M=(\mu ,11)$,  a 3-form field
$A_{MNR}$ and the gravitino, and there are a 2-brane and a
5-brane. 

The correspondence between the fields is
given by dimensional reduction from 11D supergravity~\cite{BDDN},
and in particular  the dilaton
of the 10D theory is related to the $g_{11,11}$ component of the
metric.
To  obtain directly the relation  between the 
10D metric and dilaton and the 11D metric we can
neglect for the moment $g_{\mu ,11}$, which
produces a RR one-form,
keeping only the components $g_{\mu\nu}$
of the 11D metric $g_{MN}$, $M=(\mu ,11),\mu=0,\ldots 9$, and 
the component $g_{11,11}\equiv b^2$, so that $b$ is the scale factor of the
$11^{th}$ dimension. Then a simple computation shows that
the 11-dimensional Einstein-Hilbert action becomes
\be
S=\frac{1}{2\lambda_s^8}\int d^{10}x\sqrt{-g^{(10)}}\,\, b R^{(10)}\, .
\ee
The scale factor $b$ of the  $11^{th}$ dimension looks
non-dynamical at a first sight, since it appears without derivatives. 
One should remember though that the 10-dimensional Ricci scalar 
$R^{(10)}$ contains second derivatives of the metric. The fact that $b$
is indeed dynamical is made evident by a conformal rescaling
$g^{(10)}_{\mu\nu}\ra b^{-1}g^{(10)}_{\mu\nu}$.
Defining the dilaton $\phi$ from
\be
e^{\phi}=b^{(D-3)/2},
\ee
where in our case $D=9$, one finds
\be
S=\frac{1}{2\lambda_s^8}
\int d^{10}x\sqrt{-g^{(10)}}\, e^{-\phi}\left[ R^{(10)}+
(\pa\phi )^2\right]\, ,
\ee
which is the effective action of string theory in the gravity--dilaton
sector. 

The KK field $g_{\mu ,11}$ becomes instead a RR one-form $C_{\mu}$
in type~IIA. The object which is charged under $C_{\mu}$ in type~IIA
is the D0-brane, which therefore from the 11-dimensional point of view
is coupled to $g_{\mu ,11}$ and therefore is 
a supergraviton with momentum $p_{11}=1/R_{11}$. The state with
momentum $p_{11}=n/R_{11}$, in the type~IIA language becomes a collection
of $n$  D0-branes, an interpretation made possible by the fact that $n$
D0-branes have a  marginally bound state with a mass
$n$ times the single D0-brane mass $1/(g\sqrt{\alpha '})$, and  this
result receives no perturbative or non-perturbative correction
because this state is BPS saturated.

The transition at the 10D Sugra/11D Sugra border is therefore 
different from the example discussed in the introduction, since it
does not involve a dramatic reshuffling of degrees of freedom. 
The 10D gravitons are still relevant massless modes, but they are now
embedded in 11D-gravitons with vanishing momentum in the $11^{th}$
dimension. At first sight, one might think that we can  
neglect the states with non-vanishing  momentum in the $11^{th}$
dimension (i.e. D0-branes), and that we can limit ourselves to  11D
gravitons, and look for 11D-cosmological solutions,
spatially constant in all directions including the 11$^{th}$
(for work on 11D solutions see refs.~\cite{RRfields}). 

However, the situation is more complicated and
the fields of 11-dimensional supergravity are not the only relevant
modes: in a gravitational field, 
D0-branes can become massless  and, beyond a critical field,
tachyonic.

The fact that D0-branes are  non-negligible degrees of freedom in the
11D Sugra regime is already  suggested by the computation in
Ref.~\cite{GHV} of D0-branes pair production. This can be computed
first of all in the 11D-Sugra region, considering for instance a
process of graviton-graviton scattering with production of
D0-branes. While from the 10D point of view this process is inelastic
and non-perturbative, from the 11-dimensional 
point of view it is simply an elastic scattering process
in which the  final 11D gravitons have a component of the momentum
along the $11^{th}$ dimension. The process can therefore be computed
reliably, and clearly the production of pairs of D0-branes is not
suppressed. 
In Ref.~\cite{GHV} the production rate of D0-branes has also been
computed using matrix string theory. The result, which gives an
unsuppressed rate, is found to be in
agreement with the supergravity calculation, but is expected to be
valid more in general in the M-theory regime.

In the early Universe, to produce D0-branes is not even necessary to
consider a scattering process. A time-varying gravitational field produces particles. The general mechanism is the existence of a non-trivial Bogoliubov transformation between the in and out vacuum. The 
amplification of vacuum fluctuations which takes place in any
cosmological background will produce a spectrum of 11-dimensional
supergravitons. 
{}From the 10D point of view, this means that a background
of both 10D gravitons and D0-branes  are created
from the vacuum.

The analysis is 
similar to the one performed at finite temperature to identify the 
states which   become tachyonic at the Hagedorn temperature in the 
weak coupling regime \cite{sathi,kog,aw}.
Identifying the D0-branes with gravitons with a
non-vanishing momentum along the 11$^{th}$ dimension, we need to know the
closed string spectrum in a gravitational field
at oscillator level 1, corrisponding to
gravitons, and with non-vanishing KK momentum. However,
solving the equations of motion and constraints of strings in 
generic  curved spacetime is a prohibitive task. One can try the expansion
 \cite{dvs,sv}
$X^\mu(\sigma,\tau)=x^\mu(\tau)+Y^\mu(\sigma,\tau)$, where $x^\mu$ defines 
the motion of a point-particle and 
 $Y^\mu$ describes the finite-string size effects. The physical idea
behind this expansion is that a small-enough string should follow
closely the point-particle geodetic.

At  second order in $Y^\mu$
and large values of the scale factor of the Universe $a$ 
(or  at large times), the components $Y^i$ ($i=1,\cdots,D$) satisfy a 
particularly simple equation. Defining $\chi^i=a Y^i=\sum_n \chi_n^i 
e^{i n\sigma}$,  one gets \cite{sv}
\begin{equation}
\ddot{\chi}_n^i
=\left[-n^2+\ell^2 \frac{\ddot{a}}{a}\right]\chi_n^i,
\end{equation} 
where the dot is the derivative with respect to world-sheet time, while
$\ell$ is a parameter of the background configuration around
which we are expanding. 
We can physically interprete it  as the total length of the string
and, at least for string not highly excited, we can take
$\ell\sim\lambda_s$. 

Specializing -- for simplicity -- to the case of isotropic
de Sitter geometry\footnote{We would like to stress that a period
of de Sitter expansion might not necessarily  come out as a solution 
of the 
equations of motion of the system. However, we believe that our
analysis provides some understanding 
of a generic stage of cosmological evolution 
when the curvature has a typical scale $\sim H^2$.} with constant 
$H$, the frequency modes of the string are given by \cite{dvs,sv}
\begin{equation}
\label{b}
\omega_n=\left(n^2-\ell^2 H^2\right)^{1/2}.
\end{equation}
The mass shell condition of the closed string modes can be 
written, at large times (when the world-sheet time becomes proportional to
conformal time~\cite{sv}) as  
\begin{equation}
\frac{1}{4}M^2= \frac{m^2}{4R_{11}^2}+
\frac{R_{11}^2}{(2\alpha')^2}q^2+
\frac{1}{2\alpha '}\left(N+\widetilde{N}-2c\right),
\end{equation}
where $\sqrt{2\alpha '}=\lambda_s$;
$m$ and $q$ are integers which give the  momentum and winding
in the 11$^{th}$ dimension,
$N=\sum_{n=1}^{\infty}\omega_na_{n}^{\dagger}\cdot a_n$  with
$a_n^{\mu},a_n^{\mu\dagger}$ conventionally normalized harmonic
oscillator operators for the left handed modes, and similarly
$\widetilde{N}$  for the right-handed modes, and
$N-\widetilde{N}=mq$; $c$ is the normal ordering constant.

A collection of $N$ D0-branes is a 11D-graviton with $N$ units of momentum
in the $11^{ th}$ dimension and therefore has oscillator
level $n=1$, KK quantum number $m=N$, and winding number $q=0$,
(and therefore $N=\widetilde{N}$).
Its mass is
\be\label{tac1}
M^2=\frac{N^2}{R_{11}^2}+\frac{8}{\lambda_s^2}
\left[ \left(1-\ell^2 H^2\right)^{1/2}-c\right]\, .
\ee
The constant $c$ comes from normal ordering and we will set
$c=1$ as in flat space.\footnote{Actually, here arises a subtlety
which reflects the fact that this approach to quantization in curved
background is somewhat heuristic and approximate, and cannot be taken
too literally. In flat space $c$
 is obtained evaluating $\sum_{n=1}^{\infty}n$.
This  can be done e.g. by zeta function regularization or lattice
regularization. The latter  can be performed  replacing $n$ with
$\sin (\epsilon n)/\epsilon$ and performing the sum up 
to $n_{\max}=\pi/\epsilon$~\cite{GSW}, where $\epsilon$ is the lattice
spacing, and
then subtracting the term that diverges as $1/\epsilon^2$ in the 
limit $\epsilon\ra 0$. This leaves us with $\sum_{n=1}^{\infty}n
=-1/12$, in agreement with zeta function regularization. 
In de Sitter space we should instead
evaluate  $\sum_{n=1}^{\infty}(n^2-\ell^2H^2)^{1/2}$;
for $\ell H\ll 1$, after performing a lattice
regularization and evaluating the sum~\cite{CP}, we find an additional
logaritmic divergence so that finally $c=1+6\ell^2H^2(\log L+\gamma_E)$,
with $L=\pi /\epsilon$ and $\gamma_E$ the Euler constant. 
Expanding  $(1-\ell^2 H^2)^{1/2}\simeq
1-(1/2)\ell^2H^2$ in eq.~(\ref{tac1}), we see
that this logarithmic divergence can be absorbed into a
renormalization of the total length of the string
$\ell$. Therefore the mass is finally
given by eq.~(\ref{tac1}) with $c=1$. Our conclusions on the state at
level $N=1$ are anyway
independent of the value of $c$, as long as the state with $N=0$ is
tachyonic, which is physically correct, see below.}

Eq.~(\ref{tac1}) shows various kind of instabilities. 
First of all, for large enough Hubble constant $H$, 
the frequencies $\omega_n$ become imaginary and then
the low frequency modes become unstable. This instability has been 
discussed in~\cite{sv} and has partly motivated the pre-big-bang
scenario~\cite{GV}. Furthermore,
setting $N=0$, 
we see that, even for small values of $\ell H$, the
graviton  becomes tachyonic, 
with $M^2\simeq -4(\ell^2/\lambda_s^2)H^2\lsim -4H^2$.
Physically, this is consistent with the fact that in de~Sitter space
gravitons are produced by amplification of vacuum fluctuations, with
an effective temperature $H/(2\pi )$.

Here we are interested in the appearance of further tachyonic
states, with non-zero momentum in the 11$^{th}$ dimension.
Expanding  for $\ell H\ll 1$, for $N=1$ 
a new tachyonic mode appears at
\begin{equation}
\label{c}
H\simeq \frac{1}{2}\frac{\lambda_s}{\ell} \frac{1}{R_{11}}\lsim 
 \frac{1}{2} \frac{1}{R_{11}}\equiv H_c
\end{equation}
which corresponds to the solid line of fig. 1.
Note that $\ell H_c\sim \lambda_s/(2R_{11})=1/(2g)$ so that our
expansion for $\ell H\ll 1$ is consistent in the region
$g\gg 1$. Without relying on this expansion, we read from
eq.~(\ref{tac1}) that a tachyon indeed exists, for each $N$, if
$N^2/R_{11}^2-8/\lambda_s^2<0$, or if $g>N/(2\sqrt{2})$. The first 
new tachyon, $N=1$, appears if $g>1/(2\sqrt{2})$, which is consistent with
the fact that we are working at $g>1$ (actually, we could now take
$g=1/(2\sqrt{2})$ as the more precise location of the `triple
point' in the phase diagram, fig.~1).

This instability is also manifest in the closed string 
vacuum-vacuum transition amplitude at one-loop, ${\cal A}_{\rm 1-loop}$. 
The amplitude is  given by
\be
{\cal A}_{\rm 1-loop}=iV_d\int\frac{d\tau d\bar{\tau}}{4\tau_2}
(4\pi^2\alpha '\tau_2)^{-d/2} \sum z^{N-1}
\bar{z}^{\widetilde{N}-1}\, ,
\ee
where $\tau =\tau_1+i\tau_2$, the integration is over the fundamental
region of the torus, $z=\exp (2\pi i\tau )$ 
and the sum is over the physical closed string
spectrum. The limit $\tau_2\ra\infty$ is dominated by the lightest
modes~\cite{Polbook}
\be
{\cal A}_{\rm 1-loop}=2\pi iV_d\int^{\infty}\frac{d\tau_2}{2\tau_2}
(4\pi^2\alpha '\tau_2)^{-d/2}
\sum_i\exp (-\pi\alpha' m_i^2\tau_2)\, .
\ee
and diverges if a tachyon, $m_i^2<0$, is present. 
The instability signals that new channels are opening up.
Therefore,  when the horizon  of the Universe contracts  
approaching the  critical value $H_c^{-1}$, the effective theory 
from 10-dimensional becomes 11-dimensional and this is signalled 
by the copious and unsuppressed gravitational  production of D0-branes  
carrying longitudinal  momentum $p_{11}=1/R_{11}$ and -- eventually -- states 
carrying KK momentum $p_{11}=N/R_{11}$ which are properly described as
marginally  bound
states of $N$ D0-branes.
At $H\sim H_c$ and strong coupling the Universe  enters a phase where
the energy density is dominated by the gas of branes produced at large
curvature. The production of this gas has  
clearly  a regularizing effect on the growth of the curvature
in a pre-big-bang type scenario. From
the 10D point of view,  the energy density
grows until it reaches a critical value, where it is dissipated
by generating D0-branes. Probably, inhomogeneities in the field
configuration play here an important role, as seeds for D0-branes
formation. 

{}From the 11D point of view, instead,
 above a critical value the classical field, which was constrained to live
in 10D, starts to spread over the 11$^{th}$ dimension, and its energy
density consequently drops.
When the Universe enters this
phase of D0-brane domination  the regime of growing curvature is
significantly slowed down (see also~\cite{Ram}).

The transition at high curvature and non-perturbative regime into a
new phase of gas of D0-branes that
we have just described is reminiscent of the Hagedorn transition that
takes place at finite temperature in the matrix model \cite{matrix} of
$N$ D0-branes. At low temperatures, there is a string phase where the
D0-branes are spread out to form a membrane wound around a compact
direction. In this phase the ground state must be that of a IIA string
with a specified value of the string tension. The light modes are
those of a string. At high temperatures the energy in the strings is
overwhelmingly found in the highest mode that can be excited with that
energy. However, in the matrix model there is a cut-off on the mode
number. Thus in the partition function there is a natural cut-off in
the energy integral above which the discrete nature of the string
becomes important and at this point one has to go back to the original
matrix model. 
What happens is that, at high temperatures, the D0-branes prefer 
to cluster at one point, thus the  string ceases to exist. 
One can think of this as the tendency of the string to shrink to zero 
size and disintegrate into a bunch of D0-branes.  
The light degrees of freedom are therefore the  massless modes
 which are just the fluctuations about the origin of the $N^2$ matrix 
elements. This phase transition can be 
identified with the Hagedorn transition for the 
IIA string 
constructed 
out of the $N$ D0-branes \cite{hag}.

\section{The D$p$-brane phase}\label{5}

The goal of this section is to describe the  dynamics of the system when 
the curvature of the Universe approaches values of the order of $\ell_p^{-2}$.
 To do so, we have first to study the dynamics of D$p$-branes in 
curved spacetime.

\subsection{D$p$-brane dynamics in curved background}

\label{ten}

The action governing the dynamics of a Dirichlet
$p$-brane is~\cite{Dbrane,Lei}
\be\label{S0}
S=-T_p\int d^{p+1}\xi\, e^{-\phi}\sqrt{-
\det (G_{\alpha\beta}+B_{\alpha\beta}+2\pi\alpha 'F_{\alpha\beta})
}\, ,
\ee
where $T_p^{-1}=\sqrt{\alpha '} (2\pi\sqrt{\alpha '})^p$,  
$\xi^{\alpha}$ ($\alpha=0,\ldots ,p$)
parametrize the D-brane world-volume, 
$G_{\alpha\beta}$ is the induced metric
\be\label{induced}
G_{\alpha\beta}=g_{\mu\nu}(X)\pa_{\alpha} X^{\mu}\pa_{\beta} X^{\nu},
\ee
while $B_{\alpha\beta},F_{\alpha\beta}$ 
are the pull-back of $B_{\mu\nu},F_{\mu\nu}$ to the
brane. If we consider a flat background metric, 
$g_{\mu\nu}=\eta_{\mu\nu}$, and we
assume that the extrinsic curvature of the brane is small, 
then we can separate the $D+1$ coordinates $X^{\mu}$ into 
$p+1$  longitudinal components $X^{\mu}(\xi )= \xi^{\mu}$ ($ \mu =0,\ldots ,p$)
and $(D-p)$ transverse coordinates $X^a$ (where $D$ is the appropriate number of
spatial dimensions).
The induced metric then becomes 
\be
G_{\alpha\beta}\simeq\eta_{\alpha\beta}
+\pa_{\alpha}X^a\pa_{\beta}X^a\, .
\ee
Neglecting $B_{\mu\nu}$ and expanding to second order in $\pa X$ and 
$F$ one gets~\cite{Dbrane}
\be\label{flat}
S=-\tau_p V_p
- \int d^{p+1}\xi\,\left[ \frac{\tau_p}{2}\pa_{\alpha}X^a\pa^{\alpha}X^a
+\frac{1}{4g_{YM}^2}
 F_{\alpha\beta} F^{\alpha\beta}\right]
\, ,
\ee
where $\tau_p=T_p/g$ is the brane tension, $V_p=\int d^{p+1}\xi$ is
the brane volume, 
$g_{YM}^2=(g/\sqrt{\alpha '}) (2\pi\sqrt{\alpha '})^{p-2}$
and $g=e^{\phi}$. The term $\tau_pV_p$ is the mass of the D-brane, while
the fields $A_{\alpha}$ and $X^a$ describe the D-brane fluctuations. The
 longitudinal components $A_{\alpha}$ of $A_{\mu}$ are 
governed by
a $U(1)$ gauge theory living on the brane, while the transverse
displacements $X^a$ are $(D-p)$ world-volume scalars.

We now repeat the procedure for a curved background metric
$g_{\mu\nu}$. We concentrate on the transverse coordinates $X^a$ and
we set $F_{\mu\nu}=B_{\mu\nu}=0$. The most convenient 
generalization of the notion of  transverse coordinates to the case of a
curved background is provided by Riemann normal coordinates. Given a
brane configuration $\bar{X}^{\mu}(\xi )$, arbitrary 
fluctuations transverse to
this configuration are parametrized by $(D-p)$ world-volume scalars
$\phi^a(\xi )$ as
\be\label{riemcoord}
X^{\mu}(\xi )=\bar{X}^{\mu}(\xi )+\bar{n}^{\mu}_a(\xi )\phi^a(\xi ),
\ee
where $\bar{n}_a^{\mu}(\xi )$ ($a=1,\ldots ,(D-p)$) are the $(D-p)$ 
vectors normal to the brane at the point
$\bar{X}(\xi )$. The expansion of the metric in Riemann
coordinates is~\cite{Pet}
\be
g_{\mu\nu}(X)=\bar{g}_{\mu\nu}-\frac{1}{3}\phi^a\phi^b
\bar{R}_{\mu\rho\nu\sigma} \bar{n}_a^{\rho} \bar{n}_b^{\sigma}
+{\cal O}(\phi^3)\, ,
\ee
where the overbar indicates that the quantity is evaluated at
$\bar{X}$. If the extrinsic curvature of the membrane is small, the
normal vectors are 
$\bar{n}_a^{\rho}\simeq\delta_a^{\rho}$, and therefore 
\be\label{exp}
g_{\mu\nu}(X)=\bar{g}_{\mu\nu}-\frac{1}{3}X^aX^b
\bar{R}_{\mu a\nu b}+{\cal O}(X^3).
\ee
(We now use the  notation
$X^a\equiv \phi^a$, so that from now on $X^a$ are the
fluctuations over $\bar{X}$, rather than $X^{\mu}$ for 
$\mu =a$. We also use the sign conventions 
$\eta_{\mu\nu}=(-,+,\ldots ,+), 
R^{\mu}_{\hspace*{2mm}\nu\rho\sigma}=
\pa_{\rho}\Gamma^{\mu}_{\nu\sigma}+\ldots$).
Again, we  specialize  to the de~Sitter background for 
simplicity, but our conclusions can be generalized to any specific
situation where $H^2$ is the typical curvature scale.
The metric is
\begin{equation}
ds^2=a^2(\eta )(-d\eta^2
+dx_idx^i).
\end{equation}
 The scale factor is $a(\eta )=-1/(H\eta )$, and $\eta$ is
conformal time, $\eta <0$. The Riemann tensor is given by 
$\bar{R}_{\mu\nu\rho\sigma}=H^2
(\bar{g}_{\mu\rho}\bar{g}_{\nu\sigma}-
\bar{g}_{\mu\sigma}\bar{g}_{\nu\rho})$ and the induced metric 
 in the
limit of small extrinsic curvature is then
\be\label{ind}
G_{\alpha\beta}= F_{\alpha\beta}+f_{ab}\pa_{\alpha}X^a\pa_{\beta}X^b,
\ee
where
\bees\label{H}
F_{\alpha\beta}&=&a^2\left( 1-\frac{1}{3}H^2a^2X^2 +{\cal O}(X^4)
\right)
\eta_{\alpha\beta},\\
f_{ab}&=&a^2\left[ \left( 1-\frac{1}{3}H^2a^2X^2
\right)\delta_{ab}+
\frac{1}{3}H^2a^2X_aX_b+{\cal O}(X^4)\right]\, ,
\ees
and $X^2=\sum_a X^aX^a$. Here  $X^a$ represents the   comoving coordinate and 
$X_{p}^a=a(\eta)X^a$
is the physical coordinate, which expands with the scale factor.

We are now in the position to  compute $\det G_{\alpha\beta}$. 
Retaining only terms
up to second order in $X$ and $\pa X$ we get 
\be\label{mass}
S=-\tau_p V_p
-\tau_p \int d^{p+1}\,\xi a^{p+1}\, \left[
\frac{1}{2}\eta^{\alpha\beta}\pa_{\alpha}X^a\pa_{\beta}X^a
-\frac{(p+1)a^2H^2}{6}X^aX^a\right]\, . 
\ee
In flat space the fields $X^a$  are massless, see eq.~(\ref{flat}), 
since they are the
Goldstone bosons of broken translation invariance in the transverse
directions. 
Recalling that our signature is $\eta_{\alpha\beta}=(-,+,\ldots ,+)$,
we see that in curved space the pseudo-Goldstone 
bosons of broken translation invariance, $X_p=aX$,  
get a {\it tachyonic} mass
\be\label{tac}
m^2=-\frac{(p+1)H^2}{3}\, .
\ee
The situation is very similar to the thermal tachyon which occurs at
the Hagedorn temperature \cite{sathi,kog}. In that case a condensate of
winding states forms when the periodic imaginary times becomes too
small. In the present case the tachyon forms because of the nontrivial
geometry of the spacetime.

Switching to cosmic time $t$ and expanding  $X^a (t,\vec{\xi} )
=\sum_k X^a_k(t) 
\exp (i\vec{k}\cdot\vec{\xi })$, the equation of motion for 
the comoving modes $X^a_k$ is
\be
\label{ta}
\ddot{X}^a_k +(p+2)H\dot{X}^a_k
+\left( \frac{k^2}{a^2}-\frac{p+1}{3}H^2\right)X^a_k=0\, ,
\ee
where  the dot stands for  the derivative with respect to 
cosmic time  $t$. Let us first discuss the behaviour of  the mode
$X^a_k$ in the absence of the tachyonic mass, that is dropping 
by hand the 
term $\sim (p+1)H^2/3$ in eq. (\ref{ta}). The behaviour of a mode
$X^a_k$ is very simple. As long as the physical wavelength $ak^{-1}$ of a
mode is inside the Hubble radius, 
$aX^a_k$ is an oscillating function whose amplitude
descreases as $e^{-\frac{p}{2}Ht}$. As the
physical wavelength grows it crosses
the Hubble radius and the {\it comoving} amplitude $X^a_k$  becomes
frozen, so the physical amplitude grows as $a$. Returning to the case
of interest in which the tachyonic mass is present, as long as the
physical wavelength is well inside the horizon, again the physical
amplitude  $aX^a_k$ decreases as $e^{-\frac{p}{2}Ht}$. However, as the
physical wavelength crosses the horizon, the comoving amplitude does not
remain frozen, but instead grows exponentially, so the transverse
physical amplitude
grows faster than the scale factor $a$ and therefore faster than
the physical coordinates defining the brane. 
Therefore, any  comoving transverse
fluctuations
of the D$p$-brane -- even though initially damped -- will eventually grow  
and give rise to an instability.  As the brane is stretched by the
expansion of the
Universe, transverse fluctuations grow so quickly that
the D$p$-brane cannot be
considered as 
a static object (in comoving coordinates) any longer. In fact the brane looses
its own identity as
physical transverse fluctuations grow  faster than the physical coordinates
defining the D$p$-brane.

To get further insight and to follow the system beyond the approximation
of small transverse fluctuations,  
we must compute higher order terms in the tachyon effective potential.
Let us consider the case 
of $X^a$ not necessarily small  but
$\pa_{\alpha} X^a =0$,  so that
we are interested in the mass of the brane itself, which in flat space
is the term
$\tau_pV_p$ in eq.~(\ref{flat}), rather than in the
kinetic term of the fields $X^a$. For symmetric
spaces like de Sitter space, the expansion of the metric in Riemann
coordinates can be written explicitly  at all orders in $X^a$
\cite{Pet}\footnote{Compared to eq.~(7.20) of Ref.~\cite{Pet} we have
changed the sign in front of $\sum_k$, to compensate for the fact that
Ref.~\cite{Pet} has an unusual definition
$m_{\mu\nu}=-\bar{g}_{\mu\sigma}{m^{\sigma}}_{\nu}$ while we use
$m_{\mu\nu}=\bar{g}_{\mu\sigma}{m^{\sigma}}_{\nu}$.}
\be\label{petrov}
g_{\mu\nu}=\bar{g}_{\mu\nu}-\frac{1}{2}\sum_{k=1}^{\infty}
\frac{ (-1)^k 2^{2k+2}}{(2k+2)!}m_{\mu}^{\sigma_1}
m_{\sigma_1}^{\sigma_2}\ldots m_{\sigma_{k-1}\nu}\, ,
\ee
where
\be
m_{\mu\sigma}=\bar{R}_{\mu ab\sigma}X^aX^b=
(\eta_{\mu b}\eta_{\sigma a}-\eta_{\mu\sigma}\eta_{ab})y^ay^b\, .
\ee
The last equality holds for de Sitter space and we have defined 
$y^a=HX_p^a$.
If we set $\pa_{\alpha} X^a=0$,
the induced metric 
$G_{\alpha\beta}$, for small extrinsic curvature\footnote{The
assumption of small extrinsic curvature enters when we compute
$\pa_{\alpha}X^{\mu}$ in eq.~(\ref{induced}), using the expansion 
(\ref{riemcoord}) for $X^{\mu}$, with $\mu$ a transverse index. We see
that it vanishes if both $\pa_{\alpha}\phi^a \equiv \pa_{\alpha} X^a=0$,
and $\pa_{\alpha}\bar{n}_a^{\mu}=0.$}
is simply equal to $g_{\mu\nu}$
with $\mu =\alpha ,\nu=\beta$ (recall that $\mu =(\alpha ,a)$ where
$\alpha$ runs over the value zero and the longitudinal  directions, 
while $a$ runs over the transverse directions). Eq.~(\ref{petrov})
then simplifies considerably since, if the index $\mu$ has a
`longitudinal'  value $\alpha$, then $m_{\mu}^{\sigma_1}$ is
non-vanishing only if also $\sigma_1$ has a longitudinal value
$\gamma$, and $m_{\alpha}^{\gamma}=-\delta_{\alpha}^{\gamma}y^2$, with
$y^2=y^ay^a$, and so all indices $\sigma_1,\ldots\sigma_{k-1}$ in 
eq.~(\ref{petrov}) are longitudinal. As a result, we find
$G_{\alpha\beta}=f(y)\eta_{\alpha\beta}$ with
\be\label{f}
f(y)=1-2\sum_{k=1}^{\infty}\frac{(2y)^{2k}}{(2k+2)!}=
2-\frac{\sinh^2y}{y^2}\, .
\ee
The action in the limit of small
$\pa_{\alpha} X^a$ is therefore
\be\label{S}
S=-\int d^{p+1}(a\xi )\, \tau_p   f^{\frac{p+1}{2}}(HX_{p}) \, .
\ee
The function $f(y)$ decreases monotonically, has the perturbative
expansion $f(y)=1-(1/3)y^2+\ldots$ around $y=0$  and vanishes at
$y=y_0\simeq 1.49$. Therefore, in de Sitter spacetime the ``potential" 
$f(y)$ of the tachyon field $y$
is negative  around the origin $y=0$  and the field rolls down away from it.

The existence of the critical point $y_0$ might have very interesting
implications. If
  $HX_p$ approaches the critical value, the D$p$-branes
become almost tensionless, for any value of $p$.
This  renormalization of the brane tension  is
 similar to what happens to   $(n, 1)$ strings,  
bound states of one D-string 
and $n$ fundamental strings. This system  
may be described by a D-string with a world volume 
electric field $E$ turned on. As the electric field approaches its 
critical value, 
$n$ becomes large. 
The effective tension of the $(n,1)$ string is renormalized by the
electric field, $T_{eff}=T_1 (1-E^2)$ ~\cite{GKP}. 

Our result does not
come as a surprise since the analysis performed above for small fluctuations
already suggested that the D$p$-brane violently fluctuate along
the transverse directions paying no price in energy. 
Once some D$p$-branes are formed -- as we expect at strong coupling -- 
they fluctuate along their transverse directions. Long-wavelength transverse
fluctuations grow and this growth is probably accompanied by a huge
production of light  pseudo-Goldstone bosons of the broken translation
invariance. Since the branes become tensionless, it becomes easier             
and easier for the gravitational background to produce further
massless D$p$--branes. On the other hand, when the transverse sizes
of the D$p$-branes become larger than the horizon, the branes
presumably break down and decay. Smaller branes are continuously
created and the situation is quite similar to what happens in a 
network of (cosmic) strings. One can also envisage various other
dynamical phenomena like the melting of two or more branes when they are
closer than the horizon distance.

Furthermore, we know  that D0-branes carry an halo around
them of size $\sim \ell_p$ even though they appear point-like. Therefore, 
in the context of M-theory it is  not sensible to require that the
fluctuations are smaller  then a value of order of the 11-dimensional
Planck length $\ell_p$. Even in the presence of   `minimal'
fluctuations with $\langle X^2\rangle\sim\ell_p^2$, 
at $H\sim 1/\ell_p$ there must be 
a change of regime. In fact -- if $H$ 
approaches the line labelled ``D-branes'' in fig.~1 --
$HX_p$  approaches its critical value. 
The stage of  increasing curvature should therefore  
come to an end and be replaced by a phase where the
Universe is dominated by a gas of D$p$-branes and their excitations.
Let us  observe that -- even if we have discussed D-branes
with a neglegible extrinsic curvature -- the argument can be generalized to
arbitrarily bent D-brane. For membranes, this has been done
in~\cite{memb},  where it has been shown that the extrinsic curvature
tensor $K_{\alpha\beta}$ gives a further tachyonic contribution to the
mass squared, $m^2\sim~-K_{\alpha\beta}K^{\alpha\beta}<0$.

It is 
 intriguing that the  limiting value of the curvature is consistent with
considerations based on the holographic principle as we  now discuss.

\subsection{The holographic principle and cosmology}

Recent developments in black hole physics  and in string
theory have inspired  the so-called  holographic
 principle \cite{hooft,sus}.
It requires that the degrees of freedom of a given spatial 
region live not in the interior as in ordinary quantum field theory, 
but on the surface of the region. 
The cosmological version of the holographic principle~\cite{hc}
 requires that the degrees of freedom of a spatial volume of coordinate size 
the horizon length $\ell_H$ should not exceed the area of the horizon in 
Planck units. If one accepts  this argument, there are several 
implications  when  applying it  to 11-dimensional 
theories\footnote{In the context of pre-big-bang cosmology
the holographic principle has been recently discussed in
 Refs. \cite{Rey2}.}. 

The holographic principle
 leads us to conjecture that, no matter what is responsible 
for the appearance of branes as fundamental degrees of freedom 
in the Universe and no matter what   is the dynamics of the system after
it enters the phase of brane domination,   {\it the horizon length $\ell_H$
  cannot be smaller than the 11-dimensional Planck length $\ell_p$}. 
Indeed, let us call the spatial direction $x^{11}$ longitudinal and 
the nine remaining dimensional coordinates 
transverse. 
Suppose that the  transverse area is occupied by a system of  longitudinal 
momentum $p_{11}$. Being the number of degrees of freedom  certainly larger 
than 
the number $(p_{11}R_{11})$ of KK modes excited  in the 11$^{th}$ 
dimension, the holography principle   imposes the following bound
\begin{equation}
\label{ho}
\left(\frac{\ell_H}{\ell_p}\right)^9>p_{11}R_{11}>1,
\end{equation}
or 
\begin{equation}
\label{hol}
\ell_H>\ell_p.
\end{equation}
The last inequality of eq. (\ref{ho})  
has been obtained using the fact that   D0-branes
carry a longitudinal momentum equal to $1/R_{11}$. 

To motivate even further our conjecture (\ref{hol}), we can give it the 
following  simple interpretation. As we have seen above, in the regime of 
growing large curvature the horizon length $\ell_H$ is contracting 
and the Universe becomes populated by a gas of branes. As in this gas two 
D0-branes move way from each other, they transfer continuosly their energy 
to any open strings that happen to stretch between them. A virtual pair of 
open strings can thus materialize from the vacuum and slow down, or even 
stop completely the motions of the branes (see Bachas in \cite{Dbrane}).
This phenomenon is similar to the more familiar pair production in 
a background  electric field \cite{BP} and to the appearance of an 
instability in the D-brane self-energy at finite temperature and weak 
coupling regime at the Hagedorn temperature \cite{vaz}.

We believe that the onset of the dissipation  puts a lower limit on the 
distance scales probed by the scattering of D0-branes and it turns out that 
the dynamical size of D0-branes is comparable to the inverse cubic root of 
the membrane tension, {\it i.e.}
 to the $11^{th}$-dimensional Planck scale of M-theory! It is therefore clear
 that the  horizon length  cannot contract to 
values smaller than $\ell_p$, since at this scale the dissipation sets in, 
slowing down the growth of the curvature and -- ultimately -- stopping it.

\section{Discussion and conclusions}

We now discuss the relevance of our analysis for different
cosmological issues. Our considerations can be applied for instance
to the pre-big-bang scenario~\cite{GV}.  The 
 cosmological evolution  starts at $t\ra -\infty$ in the weak
coupling, low curvature region, {\it i.e.} near the origin in fig.~1, where 
we can use the low energy effective 
action of string theory, neglecting $\alpha '$ and loop
corrections. In the string frame ({\it i.e.} when the Einsten term 
in the action has a  factor $e^{-\phi}$ in front)
the general homogeneous vacuum solution has the form~\cite{GV} 
\be\label{pre}
a_i(t)\sim (-t)^{\alpha_i}\, ,\hspace*{10mm}
\phi (t)\sim \phi_0-\gamma\log (-t)\hspace*{10mm}
(-\infty <t<0),
\ee
(where $a_i(t)$ are the FRW scale factors)
which can also be generalized to the non-homogeneos case~\cite{kasner}.
The constants $\alpha_i, \gamma$ satisfy the conditions
\be
\sum_{i=1}^D \alpha_i^2=1\, ,\hspace{10mm}
\gamma +\sum_{i=1}^D\alpha_i =1\, .
\ee
Setting $\gamma =0$ one recovers the well known Kasner solution of
general relativity. In the isotropic case, the Kasner condition fixes
$\alpha =\pm 1/\sqrt{D},\gamma =1\mp\sqrt{D}$.
The solution with the lower sign is superinflationary,
$\dot{H}>0$, with a growing dilaton, and formally reaches a
singularity as $t\ra 0^-$. In the $(H,g)$ plane, these solutions read
\be
H=\frac{1}{\sqrt{D}}\left(\frac{g}{g_0}\right)^{\frac{1}{1+\sqrt{D}}}\, ,
\ee
where $g_0$ is the initial coupling.
For $D=9$, it 
has the form $H\sim g^{1/4}$, as shown in fig.~2, for two different
values of the initial conditions.

\begin{figure}[t]
\centering
\psfig{file=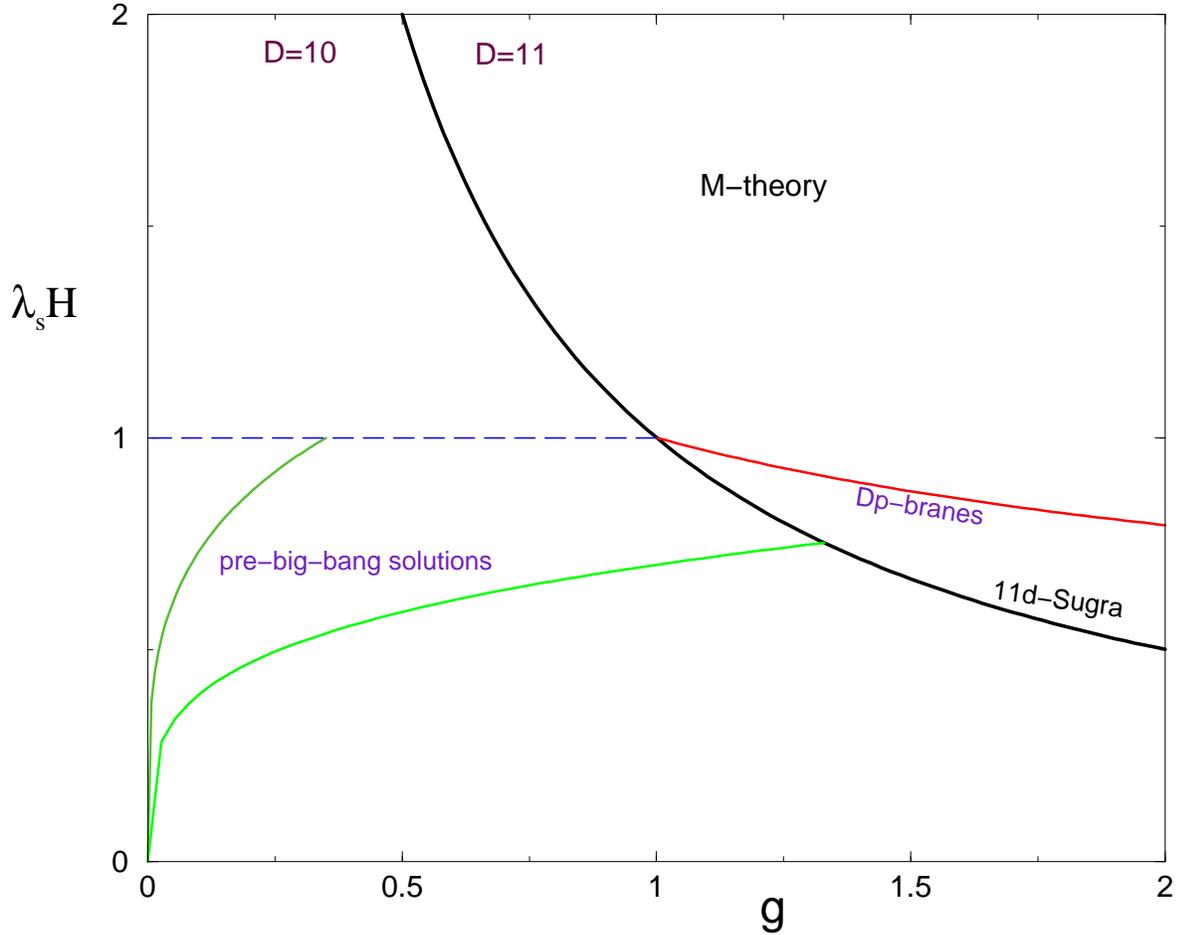,width=\linewidth,angle=270}
\caption{The phase diagram of M-theory and the
lowest order pre-big-bang solutions $H\sim g^{1/4}$, for two different
values of the initial conditions.}
\end{figure}

The classical solutions displayed in fig.~2 are only valid close to
the origin of the $(H,g)$ plane. As $\lambda_sH$ or $g$ approach one, 
the form of the solutions is determined by 
$\alpha '$ and loop corrections~\cite{corr1,corr2}. Various arguments,
based both on the study of perturbative $\alpha '$ corrections and on
the backreaction due to massive string modes~\cite{corr2}, indicate
that the evolution on the 10D side and weak coupling 
cannot go beyond the region $\lambda_sH\sim 1$, see fig.~2. If the
solution reaches this regime, the subsequent evolution typically has
$g$ increasing and $H$ constant or decreasing.
On the strong coupling side of the 10D region, on the other hand, our
results suggest that, even if loop corrections should not stop 
$H$ and $g$ from growing,
a change of regime will take place at the latest 
along the curve $\lambda_sH=1/g$,
because of the huge production of relativistic D0-branes.  
A new radiation-dominated 
regime is expected to take over when the energy density of
the light D0-branes becomes comparable to the critical energy density
for closing the Universe. Similar considerations
have been put forward in ref.~\cite{BMUV2}
in the context of a 4D compactification of string theory at weak
coupling. 

Further investigations are needed to understand whether
the evolution may cross into the 11D region -- and  possibly
approach the line labelled ``D$p$-branes''.
If so, we think  $H=1/\ell_p$ provides the largest value probed during the
evolution. 

Our results may be also relevant for the cosmology
of the so-called brane world scenario.  
It has recently been realized that the traditional
connection in perturbative  string theory
between the string scale and the Planck mass,
$M_s= \frac{1}{2} \alpha_{GUT}^{1/2} M_{p}$ does not  need
 to be applied to all string vacua
\cite{witten,lykken,untev,bajogut,ddgr,
shiutye,bachas,kakutye,benakli,bendav}. 
A possibility  that has attracted recently
much attention is lowering the
string scale   down to the weak scale, $M_s\sim $ TeV, thus giving
chances of  testing string theory at accelerators. In these
scenarios, branes play a fundamental role. Indeed, in type I string
theory
the gravitational sector consists of closed strings propagating in the
higher-dimensional bulk, while  matter fields consist of open
 strings living on D5- and  D9-branes. Even though reconciling
this scheme with the tight
constraints from cosmology seems particularly challenging, it is clear
that a complete analysis of the cosmology of the TeV-superstring scenario
cannot be performed  neglecting the presence of fundamental objects such as
D-branes. For instance, in the brane world scheme, 
inflation may automatically
take place in the pre-big-bang era and be  driven by the
dilaton field of the closed sector of type I string theory. As we
pointed out,  D-branes represent  a key ingredient for 
understanding  how the singularity is avoided and how the Universe
enters a radiation-dominated regime. Fundamental issues
such as the  
 determination of the 
``reheating'' temperature of the Universe after the stringy
phase, 
 the stabilization of the dilaton field and of the  radii of the extra
dimensions may be thoroughlly addressed only including D-branes in the
whole picture.

\vskip 0.5cm

{\bf Acknowledgements}

We would like to thank S. Dimopoulos, G.F. Giudice, 
A. Gosh, S. Gukov   and 
especially G. Veneziano for useful conversations. 
MM thanks the Theory Division of CERN, where part of this work was done,
for the kind hospitality. AR  would like to thank the International School for
Advanced Studies, Trieste, and the  Department of Physics, University of
Pisa,  where part of this work was done, for the kind hospitality.

\def\pl#1#2#3{{\it Phys. Lett. }{\bf B#1~}(19#2)~#3}
\def\prl#1#2#3{{\it Phys. Rev. Lett. }{\bf #1~}(19#2)~#3}
\def\rmp#1#2#3{{\it Rev. Mod. Phys. }{\bf #1~}(19#2)~#3}
\def\prep#1#2#3{{\it Phys. Rep. }{\bf #1~}(19#2)~#3}
\def\pr#1#2#3{{\it Phys. Rev. }{\bf D#1~}(19#2)~#3}
\def\np#1#2#3{{\it Nucl. Phys. }{\bf B#1~}(19#2)~#3}
\def\mpl#1#2#3{{\it Mod. Phys. Lett. }{\bf #1~}(19#2)~#3}
\def\ptp#1#2#3{{\it Prog. Theor. Phys. }{\bf #1~}(19#2)~#3}


\begin{thebibliography}{99}
\bibitem{SW} N. Seiberg and E. Witten, Nucl. Phys. B426 (1994) 19; 
{\it ibid.} B431 (1994) 484.
\bibitem{Dbrane} See for example: J. Polchinski, {\it TASI
lectures on D-branes}, hep-th/9611050;\\
C. Bachas, {\it Lectures on D-branes}, hep-th/9806199;\\
W. Taylor, {\it Lectures on D-branes, Gauge Theory and M(atrices)},
hep-th/9801182.
\bibitem{Str} A. Strominger, Nucl. Phys. B451 (1995) 96.
\bibitem{OV}H. Ooguri and C. Vafa, Phys. Rev. Lett. 77 (1996) 3296.
\bibitem{GV} G. Veneziano, Phys. Lett. B265 (1991) 287;
M. Gasperini and G. Veneziano, Astropart. Phys. 1 (1993)
317; Mod. Phys. Lett. A8 (1993) 3701; Phys. Rev. D50 (1994) 2519.
An up-to-date collection of references on string cosmology can be found
at http://www.to.infn.it/teorici/gasperini/.
\bibitem{corr1}
R. Brustein and G. Veneziano, Phys. Lett. B329
(1994) 429;
N. Kaloper, R. Madden and K.~Olive, Nucl. Phys. B452
(1995) 677;
I. Antoniadis, J. Rizos and K.~Tamvakis, Nucl. Phys. B415 (1994) 497.
\bibitem{corr2}
M. Gasperini, M.~Maggiore and G.~Veneziano, Nucl. Phys. B494 (1997)
315;
R. Brustein and R. Madden, Phys. Lett. B410 (1997) 110;
Phys. Rev. D57 (1998) 71;
S.J.-Rey, Phys. Rev. Lett. 77 (1996) 1929, and hep-th/9607148;
M.~Maggiore, Nucl. Phys. B525 (1998) 413;
S.~Foffa, M.~Maggiore and R.~Sturani, gr-qc/9804077, Phys. Rev.~D, to
appear.
\bibitem{Pol} J. Polchinski, Phys. Rev. Lett. 75 (1995) 4724.
\bibitem{DKPS} M. Douglas, D. Kabat, P.~Pouliot and S.~Shenker,
Nucl. Phys. {B485} (1997) 85.
\bibitem{LW} F. Larsen and F. Wilczek, Phys. Rev. D55 (1997) 4591.
\bibitem{matrix} T. Banks, W. Fischler, S.Shenker and L. Susskind, Phys. Rev.
D55 (1997) 5112.
\bibitem{GHV} S. Giddings, F. Hacquebord and H.~Verlinde, 
hep-th/9804121.
\bibitem{BDDN} E. Bergshoeff, M. De~Roo, B.~De~Wit and P.~Van
Nieuwenhuizen, Nucl. Phys. B195 (1982) 97.
\bibitem{RRfields} A. Lukas, B. Ovrut and D. Waldram,
Phys. Lett. B393 (1997) 65; Nucl. Phys. B495 (1997) 365;
H. L\"{u}, S. Mukherji, C.N. Pope and K.-W.~Xu,
Phys. Rev. D55 (1997)  7926; N. Kaloper, Phys. Rev. D55 (1997) 3394;
N. Kaloper, I. Kogan and K. Olive, Phys. Rev. D57 (1998) 7340.
\bibitem{sathi} B. Sathiapalan, Phys. Rev. D35 (1987) 3277.
\bibitem{kog} Ya.I.Kogan, JETP Lett. 45 (1987) 709;
A.A.Abrikosov,Jr. and Ya.I.Kogan, Sov. Phys. JETP  69 (1989) 235 and
Int.J.Mod.Phys. A6 (1991) 1501.
\bibitem{aw} J.~Atick and E.~Witten, Nucl. Phys. B310 (1988) 291.
\bibitem{dvs} H.J. de Vega and N. Sanchez, Phys. Lett. B197 (1987) 320.
\bibitem{sv} N. Sanchez and G. Veneziano, Nucl. Phys. B333 (1990) 253.
\bibitem{GSW} M. Green, J. Schwarz and E. Witten, {\em Superstring
theory}, Cambridge Univ. Press, cap.~11.
\bibitem{CP} The sums can be performed explicitly using the results
of appendix B of S. Caracciolo and A. Pelissetto, Phys. Rev. D58 105007.
\bibitem{Polbook} see e.g.  J. Polchinski, {\em String theory},
Cambridge Univ. Press, 1998, sect.~7.3.

\bibitem{Ram} S.K. Rama, Phys. Lett. B408 (1997) 91.
\bibitem{hag} B. Sathiapalan, Mod.Phys.Lett. A13 (1998) 2085.

\bibitem{Lei} R. Leigh, Mod. Phys. Lett. A4 (1989) 2767.
\bibitem{Pet} see for instance A.Z. Petrov, {\it Einstein spaces} (Pergamon,
Oxford, 1969).
\bibitem{GKP} S. Gukov, I. Klebanov and A. Polyakov, 
Phys. Lett. B423 (1998) 64.
\bibitem{memb} B. Carter, J. Geom. Phys. 8 (1992) 53; 
A.~Buonanno, M.~Gattobigio, M.~Maggiore, L.~Pilo and C.~Ungarelli,
Nucl. Phys. B451 (1995) 677.
\bibitem{hooft} G. `t Hooft, {\it Dimensional reduction in Quantum
Gravity}, in ``Salamfest'', pp 284-296 (World Scientific Co,
Singapore, 1993).
\bibitem{sus} L. Susskind, hep-th/9409089.
\bibitem{hc} W. Fischler and L. Susskind,  hep-th/9806039.
\bibitem{Rey2} D. Bak and S.J. Rey, hep-th/9811008; A.K. Biswas, J.
Maharana and R.K. Pradhan, hep-th/9811051. 
\bibitem{BP} C. Bachas and M. Porrati, Phys. Lett. B296 (1992) 77.


\bibitem{vaz} M. V\`{a}zquez-Mozo, Phys.Lett. B388 (1996) 494.

\bibitem{kasner} G. Veneziano, Phys. Lett. B406 (1997) 297;
A. Buonanno, K.A. Meissner, C.~Ungarelli and G. Veneziano,
Phys. Rev. D57 (1998) 2543;
A.~Buonanno, T.~Damour and G.~Veneziano, hep-th/9806230.
\bibitem{BMUV2} A. Buonanno, K.A. Meissner, C.~Ungarelli and
G. Veneziano, JHEP 01  (1998) 004.


\bibitem{witten}
E. Witten, Nucl. Phys. B471 (1996) 135.
%
\bibitem{lykken}
J.D. Lykken, hep-th/9603133 .
%

\bibitem{untev}
N. Arkani-Hamed, S. Dimopoulos and G. Dvali, hep-ph/9803315; 
I. Antoniadis, N. Arkani-Hamed, S. Dimopoulos and
G. Dvali, hep-ph/9804398;
N. Arkani-Hamed, S. Dimopoulos and J. March-Russell, hep-th/9809124.
%
\bibitem{bajogut}
K. Dienes, E. Dudas and T. Gherghetta, hep-ph/9803466;
hep-ph/9806292; hep-ph/9807522.

\bibitem{ddgr}
K. Dienes, E. Dudas, T. Gherghetta and A. Riotto, hep-ph/9809406.
%
\bibitem{shiutye}
G. Shiu and S.H. Tye, hep-th/9805157.
%
\bibitem{bachas}
C. Bachas, hep-ph/9807415.
%
\bibitem{kakutye}
Z. Kakushadze and S.H. Tye, hep-th/9809147.
%
\bibitem{benakli}
K. Benakli, hep-ph/9809582 .
%
\bibitem{bendav}
K. Benakli and S. Davidson, hep-ph/9810280 ;
D.H. Lyth, hep-ph/9810320.
%
\end{thebibliography}
\end{document}